\documentclass[letterpaper, 10 pt, conference]{ieeeconf}  

\IEEEoverridecommandlockouts                        

\usepackage{graphics} 
\usepackage{epsfig} 
\usepackage{mathptmx} 
\usepackage{times} 
\usepackage{amsmath} 
\usepackage{amssymb} 
\usepackage{subfig}
\usepackage{siunitx}

\title{\LARGE \bf
Motion Artifact Reduction In Photoplethysmography \\ For Reliable Signal Selection
}

\author{Runyu Mao$^{1}$, Mackenzie Tweardy$^{2}$, Stephan W. Wegerich$^{2}$, \\  Craig J. Goergen$^{3}$, George R. Wodicka$^{3}$ and Fengqing Zhu$^{1}$

\thanks{$^{1}$Runyu Mao and Fengqing Zhu are with the School of Electrical and Computer Engineering, Purdue University, West Lafayette, IN 47907, USA
        {\tt\small \{mao111,zhu0\}@purdue.edu}}%
\thanks{$^{2}$Mackenzie Tweardy and Stephan W. Wegerich are with PhysIQ, Chicago, IL 60606, USA
        {\tt\small \{mackenzie.tweardy,stephan.wegerich\}@physiq.com}}%
\thanks{$^{3}$Craig J. Goergen and George R. Wodicka are with the Weldon School of Biomedical Engineering, Purdue University, West Lafayette, IN 47907, USA
        {\tt\small \{cgoergen,wodicka\}@purdue.edu}}
}

\begin{document}

\maketitle
\thispagestyle{empty}
\pagestyle{empty}

\begin{abstract}
Photoplethysmography (PPG) is a non-invasive and economical technique to extract vital signs of the human body. Although it has been widely used in consumer and research grade wrist devices to track a user's physiology, the PPG signal is very sensitive to motion which can corrupt the signal's quality. 
Existing Motion Artifact (MA) reduction techniques have been developed and evaluated using either synthetic noisy signals or signals collected during high-intensity activities - both of which are difficult to generalize for real-life scenarios. Therefore, it is valuable to collect realistic PPG signals while performing Activities of Daily Living (ADL) to develop practical signal denoising and analysis methods. 
In this work, we propose an automatic pseudo clean PPG generation process for reliable PPG signal selection. For each noisy PPG segment, the corresponding pseudo clean PPG reduces the MAs and contains rich temporal details depicting cardiac features. Our experimental results show that 71\% of the pseudo clean PPG collected from ADL can be considered as high quality segment where the derived MAE of heart rate and respiration rate are 1.46 BPM and 3.93 BrPM, respectively. Therefore, our proposed method can determine the reliability of the raw noisy PPG by considering quality of the corresponding pseudo clean PPG signal.
\end{abstract}

\section{INTRODUCTION}
Photoplethysmogram (PPG) is a non-invasive signal measured by transmitting light through the skin and measuring the reflected absorption of that light from the capillaries to assess human physiological information~\cite{tamura2014wearable}. Both heartbeat and respiratory activities are recorded in PPG which can be used to derive physiological features, including vital signs of heart rate and respiration rate~\cite{allen2007photoplethysmography}.
With the increasing popularity of wearable devices, the use of a wrist-worn sensor can provide a low-cost, low-burden mechanism for collecting PPG signals for personal health monitoring.
\begin{figure}[ht]
    \begin{center}
        \includegraphics[width=0.43\textwidth]{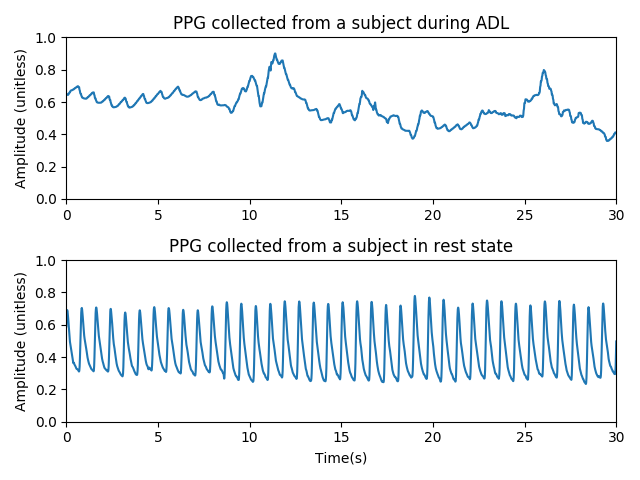}
        \caption{Comparison between PPG collected from a subject doing Activities of Daily Living (ADL) and the same subject in rest state.}
      \label{fig:active_vs_rest}
    \end{center}
\end{figure}

However, PPG signals are sensitive to various Motion Artifacts (MA). For example, PPG collected by a wrist-based sensor may be highly corrupted due to hand motion, walking, \textit{etc}~\cite{motion_artifacts}.  As shown in Figure~\ref{fig:active_vs_rest}, PPG collected from wrist devices during Activities of Daily Living (ADL) is corrupted by different types of noise. Both high frequency and low frequency noise occur non-periodically which contaminates vital features in PPG. It is difficult to determine whether the vital information has been precisely recorded in these corrupted PPG signals. On the other hand, PPG collected from the subject at rest is relatively clean and stable. Due to this reason, many existing PPG datasets~\cite{mimic,dataset_oximeter} collect reliable and clean PPG signals in lab settings, where subjects are in static states. Synthesized noises are then added to the clean PPG signal to simulate the noisy PPG signals~\cite{adaptive_1,adaptive_2,wavelet_1,wavelet_2,feature_1}. Nevertheless, both signal and noise are very different from actual PPG signals collected during ADL. 

PPG signals can also be collected where subjects are instructed to perform certain high-intensity exercises, and various methods were proposed to analyze these PPG signals, \textit{e.g.}, strong MA removal and heart rate estimation~\cite{yousefi2013motion,giannetti2012heuristic,ear_motion,intensity_wrist_2}. However, similar to the static PPG signal, these data collections were conducted in controlled settings. Thus, they struggle to accurately capture the noise due to ADL or non-period motions. 

In naturalistic settings, the quality of PPG collected during ADL is unpredictable. Some PPG segments may be highly corrupted or may not contain useful vital information. Thus, it is crucial to select reliable PPG signals collected from ADL for meaningful analysis.
To estimate the quality of PPG signals, Signal Quality Index (SQI), a binary sequence to indicate the quality of the corresponding PPG interval, has been developed based on waveform features and used in some studies~\cite{sqi_sun,sqi_li,sqi_dy}. 
Instead of focusing on the morphology of the signal, accuracy of the derived heart rate from the PPG segment is proposed to determine the PPG's reliability~\cite{real}. Although heart rate is one of the most essential vital signs for cardiac activity monitoring, crucial temporal information, \textit{e.g.}, heart rate variability, cannot be revealed. In addition, for different health applications, the quality of the PPG signal requirement may be different. For example, highly corrupted PPG signal may provide precise heart rate estimation but could yield an incorrect respiratory rate. In this case, the PPG quality can be treated as acceptable for heart rate analysis but ``unreliable" for respiratory analysis.
Therefore, a good PPG signal selection should be based on quality measurement mechanisms that contain rich temporal information and could dynamically adapt to different application requirements.

In addition, temporal information of real noisy ADL PPG could also benefit learning-based approaches for subsequent processing and analysis, \textit{e.g.}, deep neural networks based denoising. Roy~\textit{et al.}~\cite{AE} and Lee~\textit{et al.}~\cite{BAE} proposed two different auto-encoder networks for PPG denoising. Due to the scarcity of clean ADL PPG signals, synthetic noisy PPG and corresponding clean PPG signal are used for training. Although promising denoising performance are reported, it is difficult to generalize the trained model to real-life settings where noise and signal distributions may be significantly different from the training data. 

In this work, we propose an automatic, reliable ADL PPG selection framework leveraging features of electrocardiogram (ECG). Our goal is to select reliable PPG segments from collected data where both wrist PPG, chest ECG and associated features are available.
For each raw PPG segment, we produce a time-domain aligned pseudo clean PPG signal that contains rich temporal information for vital sign estimation, including heart rate and respiration rate. The quality of the raw PPG signal is determined by the vital information embedded in the pseudo clean PPG. 
Our contributions are summarized as follows:
\begin{itemize}
    \item We propose a high quality PPG selection system that leverages band-pass filter design followed by Principal Component Analysis refinement to generate pseudo clean PPG signal.
    \item Instead of providing a binary PPG signal quality indicator, our system provides pseudo clean PPG which can be used to derive vital information to assess signal quality for different applications.
    \item Our proposed PPG selection system can collect reliable PPG signals from ADL, as opposed to PPG signals with synthetic noise added.
\end{itemize}

\section{Method}\label{method}
\begin{figure*}[ht]
    \begin{center}
        \includegraphics[width=0.9\linewidth]{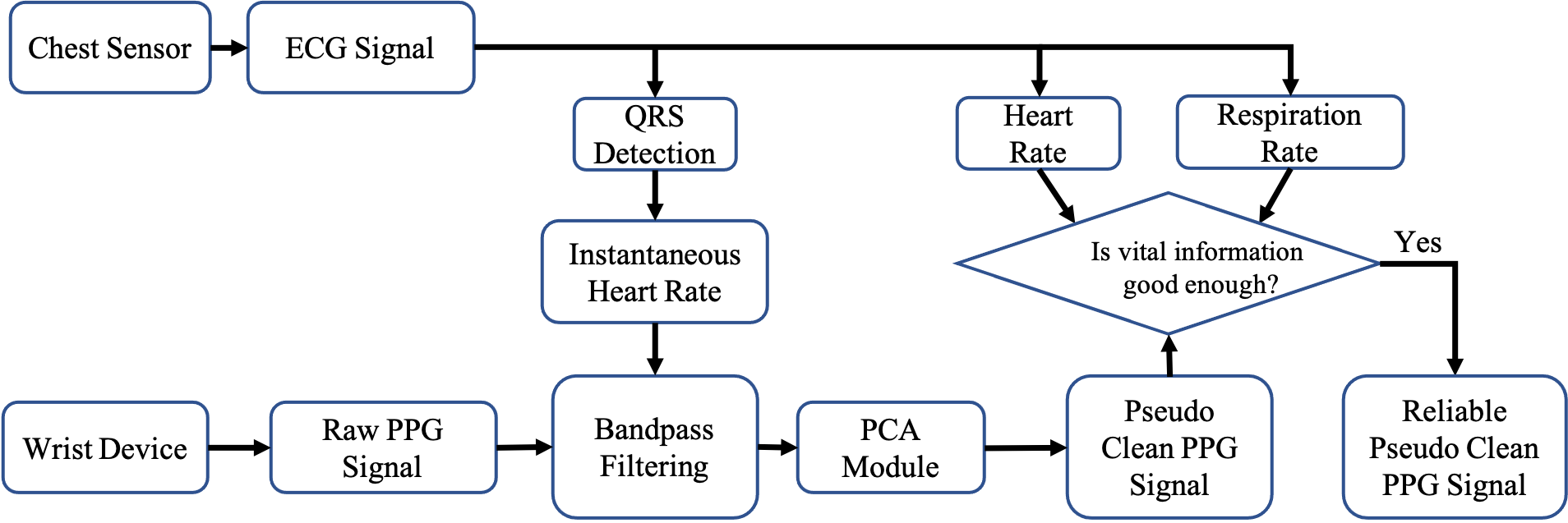}
        \caption{Overview of our proposed reliable PPG selection system.}
      \label{fig:overview}
      \vspace*{-0.2in}
    \end{center}
\end{figure*}

An overview of our proposed high quality PPG selection system is illustrated in Figure~\ref{fig:overview}. The raw single-channel PPG is collected from a wrist device, \textit{i.e.}, Samsung Galaxy Watch\textsuperscript{TM}. Additionally, simultaneous single-lead ECG and accelerometer are collected through the accelerateIQ\textsuperscript{TM} platform using the VitalPatch\textsuperscript{TM} device.
Three ECG-based features, \textit{i.e.}, heart rate, respiration rate, and QRS detection, are provided by accelerateIQ\textsuperscript{TM}. Both heart rate and respiration rate are average rates of a 1 minute ECG segment and will be treated as ground truth of each corresponding 1 minute PPG segment. The QRS detection, providing the location of each QRS complex~\cite{qrs} in ECG, will be used to derive the instantaneous heart rate for the bandpass filter design.
To measure the quality of a raw PPG segment, we generate the corresponding pseudo clean PPG signal and assess whether it contains sufficient vital information for health monitoring. The proposed pseudo clean PPG signal generation consists of a two step process. First, based on the instantaneous heart rate, we design a band-pass filter to apply to the raw PPG. Then, we use Principal Component Analysis (PCA) to further improve the signal quality. Once we obtain the pseudo clean PPG signal, we can evaluate its quality by calculating the corresponding PPG-based heart rate and respiration rate and compare them with the respective accelerateIQ\textsuperscript{TM} features. In this section, we first illustrate the signal acquisition process in Section~\ref{signal_acq}. We then describe the pseudo clean PPG generation in Section~\ref{bandpass_design} and Section~\ref{PCA}.  Finally, we discuss the process of extracting cardiovascular features from the pseudo clean PPG in Section~\ref{PPG_assessment}. 

\subsection{Signal Acquisition and Data Preparation}\label{signal_acq}
All subjects were asked to wear a Samsung's Galaxy Watch\textsuperscript{TM} on his/her non-dominant hand and VitalConnect's Vitalpatch\textsuperscript{TM} on his/her chest, simultaneously over a 7 day period. In total, 8 subjects ($5$ male and $3$ female) participated. Participants ranged in age from $27$ to $60$, with a mean age of $42.875$. Using the accelerateIQ\textsuperscript{TM} platform, QRS detection and average one minute heart rate and respiratory rate were extracted from the chest worn device. As the chest-based extracted vital signs are FDA-approved, both chest heart rate and respiration rate served as our ground truth in this work. The QRS detection records the time-points of each QRS complex~\cite{qrs} in ECG. The PPG signal was sampled at 25 Hz and all data were segmented into 1 minute intervals for analysis. The experimental procedures involving human subjects described in this paper were approved by the Institutional Review Board.

\subsection{Band-Pass Filter Design}\label{bandpass_design}
The chest patch sensor collects vital signs in a much robust manner compare with the wrist devices. Based on the detected QRS complex in chest ECG, the pauses between heartbeats (R-R interval) is recorded and used for instantaneous heart rate estimation. We select the 1 minute PPG signal and find the corresponding instantaneous heart rate range derived from the chest device. The passband of bandpass filter is selected based on the range of instantaneous heart rate so that all heartbeat activity could be preserved. Therefore, the filter should remove the MA out of the cardiac activity frequency band.

To avoid the ripple response of the digital filter, we implemented a fifth order Butterworth filter, which has a flat frequency response in the passband~\cite{IIR}. Based on the instantaneous heart rate derived from the ECG signal, a low-pass and a high-pass digital filter are designed and cascaded for narrow passband filtering. The filters process the input PPG signal in both forward and reverse directions to preserve time-domain alignment.

\subsection{Principal Component Analysis}\label{PCA}
\begin{figure}[ht]
    \begin{center}
        \includegraphics[width=0.46\textwidth]{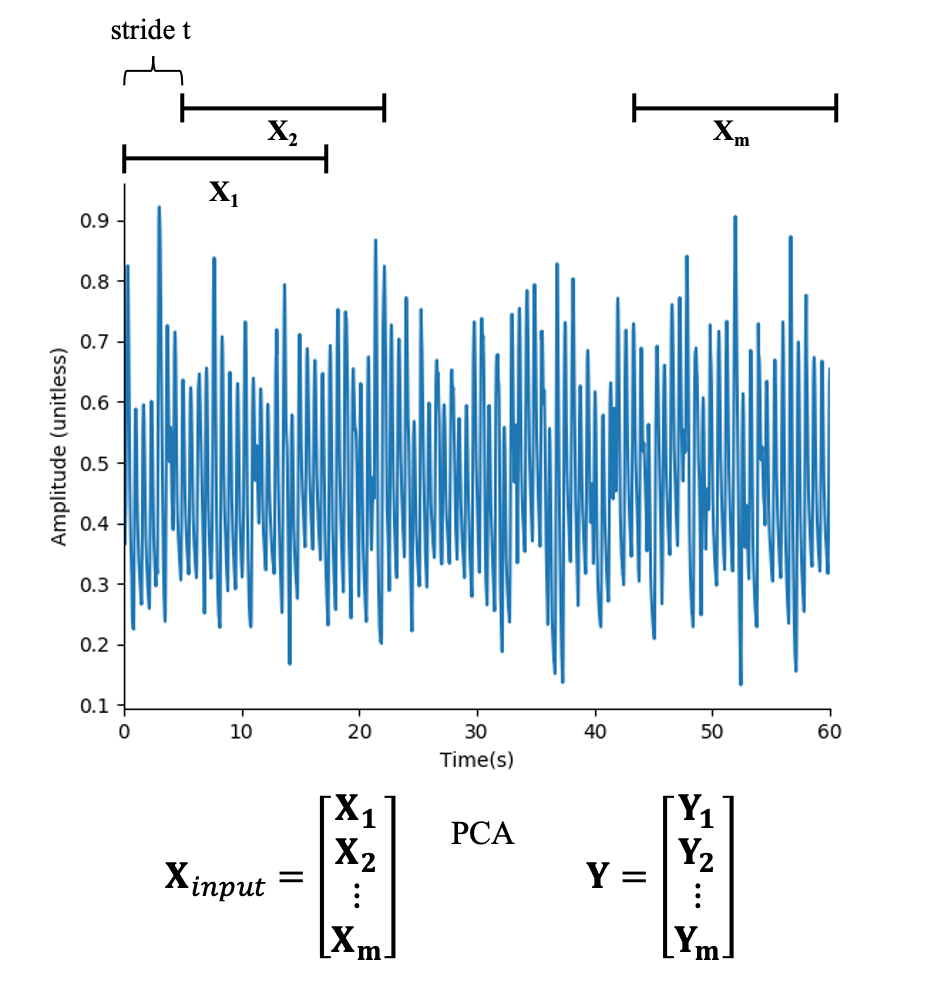}
        \caption{An example of Principal Component Analysis (PCA) on single-channel PPG.}
      \label{fig:pca}
    \end{center}
\end{figure}

To further improve the quality of filtered PPG signal, we implemented Principal Component Analysis (PCA) to remove additional noisy components. 
As shown in Figure~\ref{fig:pca}, we use a fixed $n$ length sliding window in stride of $t$ on the 1 minute filtered PPG signal,  sampled $m$ overlapped PPG observations $X_{i}$, and stacked them together as our input matrix $X_{input}$. Based on Singular Value Decomposition (SVD)~\cite{SVD}, the $X_{input}$ can at most be decomposed into $min(m,n)$ orthogonal basis.
The goal of PCA is to further reduce the MAs by selecting a subset of the basis. We select $p$ principal components and reproject them back into the input space for reconstruction. The reconstructed matrix $Y$ is also $m \times n$ that contains $m$ overlapped PPG observations corresponding to $X_{input}$. For overlapped regions in the reconstructed signal, we take the average and apply a Gaussian filter for waveform smoothing. The final output is a flattened 1 minute signal for quality assessment.

\subsection{PPG Assessment Using Heartbeat and Respiratory Activities Information}\label{PPG_assessment}
In order to evaluate the cardio-respiratory features embedded in the pseudo clean PPG signal, we derive the average heart rate and respiration rate of the PPG segment and compare to the ground truth chest-derived features.
A peak detection method based on Sobel filter~\cite{sobel} is adopted to obtain the cycle of heartbeat in pseudo clean PPG and find the corresponding heart rate value. We calculate the Mean Absolute Error (MAE) between the estimation and the reference rate from the chest features, which can be fomulated as:
\begin{equation} 
\label{eq:MAE}
MAE = \frac{\sum_{i=1}^{n}|y_{i}-x_{i}|}{n}
\end{equation}
where $y_{i}$ is reference heart/respiration rate and $x_{i}$ is our estimated heart/respiration rate. The $n$ represents the total number of segments we evaluated.
The respiratory signal is usually calculated by the temporal information in the PPG. Three kinds of Respiratory Modulation are commonly used: baseline wander (BW), amplitude modulation (AM), and frequency modulation (FM)~\cite{resp_method}. In our implementation, we use the AM and FM for respiration rate estimation since the bandpass filter removes the baseline wander.

AM of the PPG is caused by reduced stroke volume during inhalation due to changes in intrathoracic pressure, reducing pulse amplitude~\cite{AM}. FM, on the other hand, is the manifestation of the spontaneous increase in heart rate during inspiration and decreases during exhalation, known as Respiratory Sinus Arrhythmia (RSA)~\cite{FM}. Therefore, the accuracy of respiration rate estimation based on AM and FM strongly relies on the temporal information in the pseudo clean PPG.
For AM, based on the peak detection result, we extract the maximum intensity of the PPG pulses. This intensity trend sequence is resampled into an even 4-HZ sequence for spectrum analysis based on the Fast Fourier Transform~\cite{FFT}. The maximum frequency content within the Respiration Rate (RR) frequency range is selected as the RR. For FM, the peak detection for heart rate estimation provides the timepoints of each heartbeat and can be used to determine the beat interval and convert to tachograms. We resample the tachograms into an even 4-Hz grid and find the maximum power frequency within the RR frequency range in the spectrum.
The normal RR range is between 12-30 Breath/Min. (BrPM)~\cite{breath_rate}. We set the frequencies of interest to 10-50 BrPM to include all possible rates. We take the mean value to fuse the RR estimations from these two modulations.

\section{Experimental Results}
The PPG signal is collected from the Samsung Galaxy Watch\textsuperscript{TM}. We also simultaneously collect chest ECG signal from the VitalPatch\textsuperscript{TM}. Three ECG-based features, \textit{i.e.}, heart rate, respiration rate, and QRS detection, are provided by accelerateIQ\textsuperscript{TM}. The accelerateIQ\textsuperscript{TM} platform also removes the chest ECG signal during wrist device charging period and synchronized wrist PPG and chest ECG signal. A total of 55,093 samples were collected from 8 subjects. Each sample contains 1 minute segment of wrist PPG, the corresponding QRS complex timepoints, reference 1 minute average heart rate, and reference 1 minute average respiration rate. The raw PPG goes through the bandpass filter, whose passband is dynamically determined by the instantaneous heart rate derived from QRS complex timepoints and then fed to the PCA module. In the PCA module, the 1 minute 25 Hz PPG signal is a 1,500-length sequence, and we use a 400-length sliding window in stride of 5 to segment 220 16-s PPG observations and stack them together to perform PCA. We empirically select the first 30 components, which provides the best correlation coefficient between pseudo clean PPG heart rate and reference heart rate for our dataset and  construct the pseudo clean PPG signal.

\subsection{Pseudo Clean PPG Evaluation}
\begin{figure}[ht]
    \begin{center}
        \includegraphics[width=0.43\textwidth]{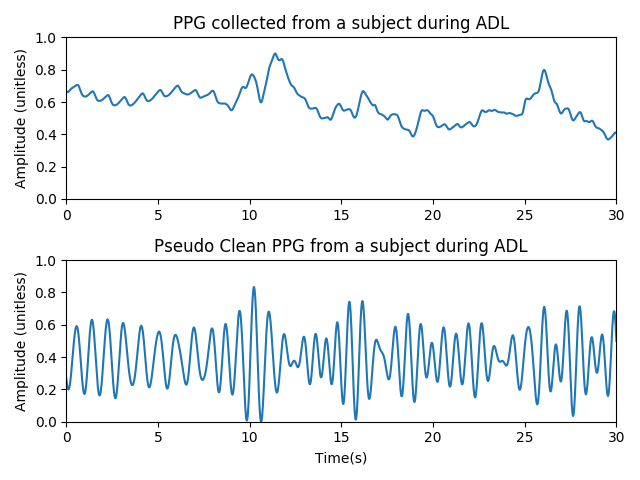}
        \vspace*{-0.05in}
        \caption{Comparison between raw PPG collected from a subject doing Activities of Daily Living (ADL) and the corresponding pseudo clean PPG generated by our method.}
      \label{fig:denoise_comparison}
    \end{center}
\end{figure}

\begin{figure}[ht]
    \centering
    \includegraphics[width=0.4\textwidth]{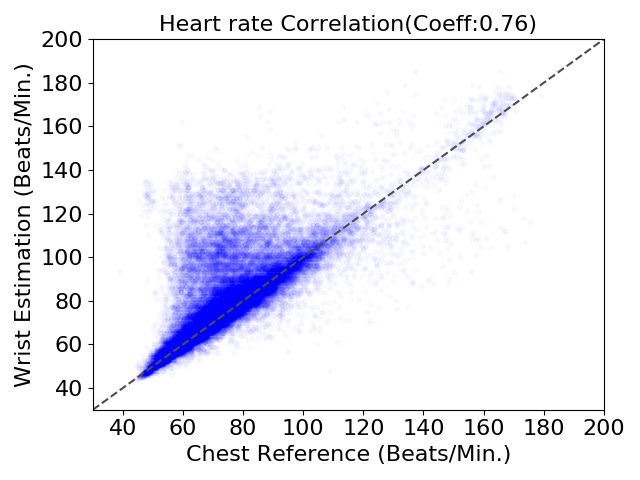}
        \caption{Pearson correlation plot of heart rate estimated from pseudo clean PPG and the chest reference rate in Beats/Min.}
      \label{fig:result_a}
\end{figure}

\begin{figure}[ht]
    \centering
    \includegraphics[width=0.4\textwidth]{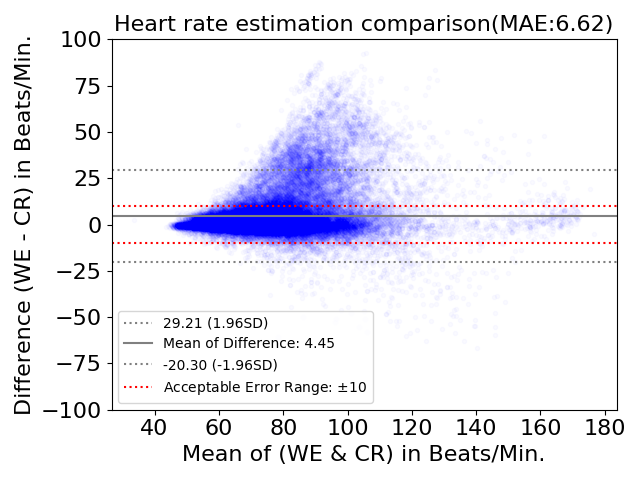}
        \caption{Bland–Altman plot of Wrist Estimated (WE) heart rate and the Chest Reference (CR) heart rate in Beats/Min. WE heart rate is derived from the pseudo clean PPG.}
      \label{fig:result_b}
\end{figure}

As shown in Figure~\ref{fig:denoise_comparison}, our method could generate a pseudo clean PPG corresponding to the noisy wrist ADL PPG signal. For better visualization, we normalize both 1 minute signals and plot the 30-s interval signals.
We first evaluate the estimated average heart rate of all the pseudo clean PPG signals and compare with the reference rates from the chest sensor. As shown in Figure~\ref{fig:result_a}, although there are some unreliable PPG segments, the Pearson correlation coefficient of $0.76$ indicates that there is a strong linear relationship between the reference heart rates and our estimated pulse rates. We also generate the Bland-Altman plot in Figure~\ref{fig:result_b}. We observe that the majority of the data points ($\sim82.49\%$) are within an general acceptable error of 10 Beats/Min. (BPM)~\cite{10bpm_1,10bpm_2}. Both plots show that accurate heart rate can be derived from most wrist PPG segments successfully. However, the data points with large error in plots also indicate that the wrist PPG collection contains the unreliable PPG segments.

To further analyze the distribution of the quality of the collected PPG, we split the PPG signals into 4 different quality groups based on the heart rate estimation. 
If the absolute error of the 1 minute average heart rate estimation is within 1 BPM, it is level-1 high quality PPG segments. The level-2 high quality PPG portion corresponds to heart rate estimation error in the range of $(1,3]$ BPM. The error range of $(3,5]$ BPM indicates level-3 high quality PPG segments. We also consider heart rate estimation error larger than $5$ BPM as low quality data.
As shown in Table \ref{table_quality}, for each group we calculate the MAE for both heart rate and respiration rate, in BPM and BrPM, based on the methods described in Section~\ref{PPG_assessment} for each group and also include the portion (\%) of PPG signals in each group to analyze the distribution.

Our results show that the pseudo clean PPG segments that yield more accurate heart rate estimation also contain higher quality temporal information to generate better respiration rate estimation. If we set the heart rate error less than 5 BPM as the threshold to collect reliable PPG segments, combining the first three group results $70.99\%$ PPG to be considered as high quality segments. In this case, the average MAE of heart rate and respiration rate are $1.46$ BPM and $3.93$ BrPM, respectively. For the low quality group, although the heart rate MAE $19.25$ BPM indicates that the pulse rate information of many PPG segments are highly corrupted, the MAE of respiration rate is still within $6$ BrPM, which means the low quality group also contains PPG segments with valid respiration rate information. If we use the respiration rate MAE as criterion for reliable PPG selection, many PPG segments in the low quality group could also be used. 

\begin{table}[h]
\caption{Heart Rate (HR) and Respiration Rate (RR) Estimation Analysis and Distributions of Different Quality PPG Segments (BPM:Beats/Min., BrPM:Breaths/Min.): Level-1 High Quality: HR error $\leq$ 1 BPM; Level-2 High Quality: 1 $<$ HR error $\leq$ 3 BPM; Level-3 High Quality: 3 $<$ HR error $\leq$ 5 BPM; Low Quality: HR error $>$ 5 BPM.}
\label{table_quality}
\begin{center}
\begin{tabular}{|c|S[table-format = 2.2]|c|c|}
\hline
\multicolumn{1}{|c|} {Quality Group}& {HR MAE}& {RR MAE}& {Portion}\\
\multicolumn{1}{|c|} {~}& {(BPM)}& {(BrPM)}& {\%}\\
\hline
Level-1 High Quality &0.46 &3.80&34.58\%\\ 
\hline
Level-2 High Quality &1.80 &3.95&25.75\%\\
\hline
Level-3 High Quality &3.89&4.32&10.66\%\\
\hline
Low Quality &19.25&5.45&29.01\%\\
\hline

\end{tabular}
\end{center}
\end{table}

\begin{figure}[ht]
    \centering
    \subfloat[Heart rate distribution of different quality groups]{\includegraphics[width=0.47\linewidth]{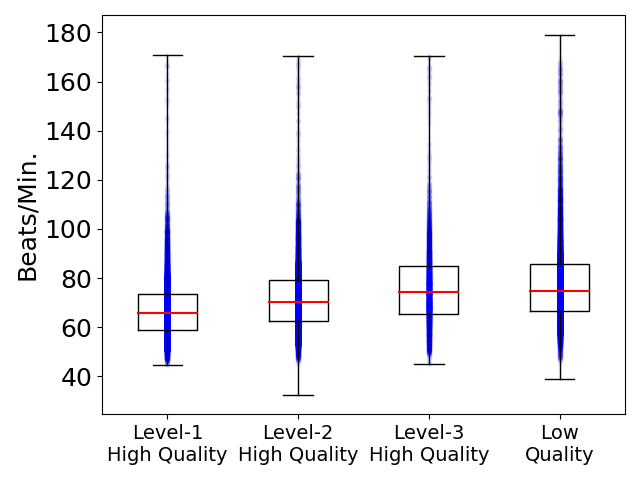}
      \label{fig:dist_heart}}
    \hspace*{0.001in}
    \subfloat[Zoom-in version of quartile values]{\includegraphics[width=0.47\linewidth]{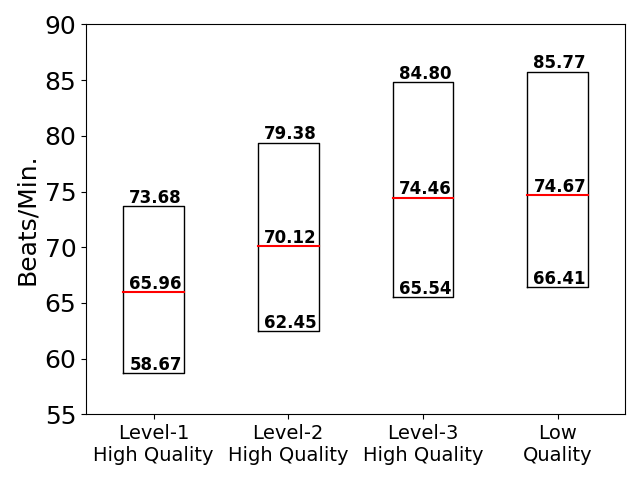}
      \label{fig:dist_heart_zmin}}
      \caption{Whisker plots depict the distributions of ground truth heart rate of each PPG group. The boxes indicate the first quartile (25\%) and third quartile (75\%) values of the heart rate and the horizontal red lines indicate the median values.}
     \label{fig:dist_H}
\end{figure}

\begin{figure}[ht]
    \centering
    \subfloat[Respiration rate distribution of different quality groups]{\includegraphics[width=0.47\linewidth]{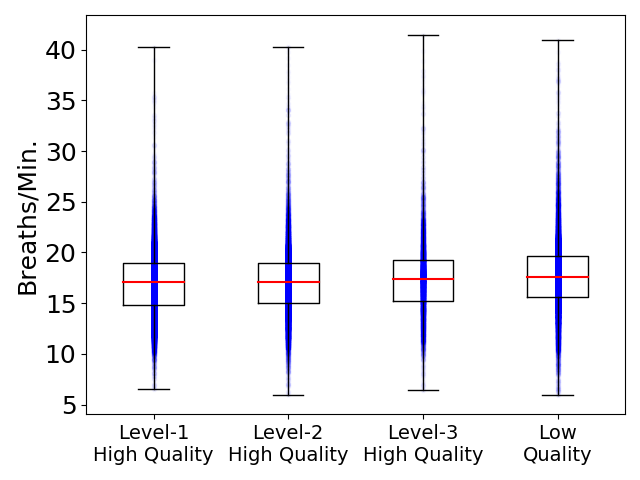}
      \label{fig:dist_resp}}
    \hspace*{0.001in}
    \subfloat[Zoom-in version of quartile values ]{\includegraphics[width=0.47\linewidth]{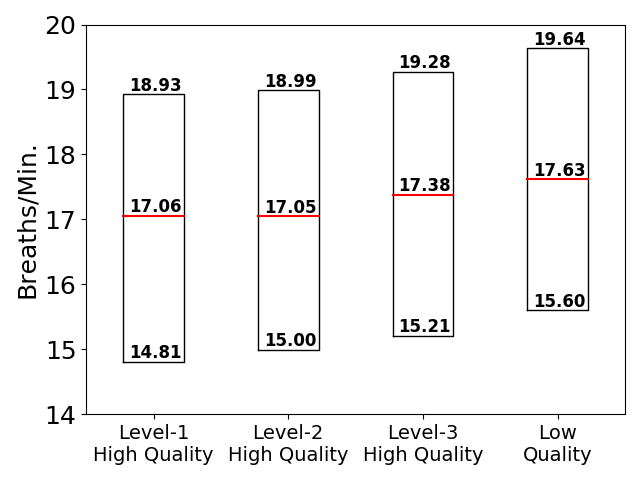}
      \label{fig:dist_resp_zmin}}
      \caption{Whisker plots depict the distributions of ground truth respiration rate of each PPG group. The boxes indicate the first quartile (25\%) and third quartile (75\%) values of the respiration rate and the horizontal red lines indicate the median values.}
     \label{fig:dist_R}
\end{figure}

For each quality group, we also include the whisker plot to show the ground truth heart rate and respiration rate distribution. As shown in Figure~\ref{fig:dist_heart} and Figure~\ref{fig:dist_resp}, the box extends from the lower quartile ($25\%$) to upper quartile ($75\%$) values of the rates, the red lines indicate the median values. Two caps show the minimum and maximum values of the heart/respiration rate of each group. We also plot the blue scatters to visualize the heart rate and respiration rate distributions. Figure~\ref{fig:dist_heart_zmin} and Figure~\ref{fig:dist_resp_zmin} show the zoom-in portions of the whisker plots to better visualize the quartile values. Although the overall distributions of both heart rate and respiration rate slightly increase as the quality decrease, both figures show that none of the quality group biases towards a specific range of heart rate or respiration rate. 

\subsection{Discussion}
Collecting reliable PPG in naturalistic setting, \textit{e.g.}, during ADLs, is valuable for health monitoring. Therefore, how to remove unreliable PPG is an important factor to consider during PPG signal selection. In this work, we propose a bandpass filter design and additional PCA refinement for pseudo clean PPG generation. The passband, as determined by instantaneous heart rate, provides the narrowband and removes MAs in other frequency range. The PCA module further removes small unrelated components of the filtered signal so that the most valuable physiological information is preserved in the pseudo clean PPG. Due to the flat frequency response of our IIR filter design, temporal details such as the amplitude of peaks are also well preserved and can be used for respiration rate estimation by the AM and FM based methods.
Our generated pseudo clean PPG can adapt to different study purposes for high quality PPG selection. For example, commercial purpose health monitoring would treat the heart rate error within 10 BPM as high quality PPG. 
In this implementation, we select heart rate error less than 5 BPM as high quality PPG and still get 71\% of collected PPG segments qualified.
For certain clinical studies, a stricter standard may be applied which requires us to consider the temporal details, \textit{e.g.}, heart rate variability (HRV) and respiration rate information, of the pseudo clean PPG to determine the selection criteria. We will explore this in our future work. In addition, our pseudo clean PPG and raw noisy PPG are aligned in the time domain because of (1) the band-pass filter processes the input PPG signal in both forward and reverse directions to eliminate the phase distortion, and (2) the PCA module is a linear transformation which does not influence the phase response. Therefore, the pseudo clean PPG selected from reliable raw PPG could be used in training-based methods for subsequent processing and analysis. 

\section{Conclusion}
In this paper, we present a novel framework to select reliable wrist ADL PPG based on the cardio-respiratory features derived from chest ECG.
The pseudo clean PPG is generated for each 1 minute raw PPG segment and embeds vital information to assess the quality of original raw PPG.
We designed bandpass filters that are guided by ECG-based features with additional PCA refinement. 
Our experimental results show that our pseudo clean PPG not only can be used for heart rate estimation but also contains rich temporal information for respiration rate prediction.
Based on the vital information quality of pseudo clean PPG and different application requirements, we can acquire reliable PPG segments for ADL.
Since the PPG is extracted from subjects in the real-life settings, the natural MA recorded in raw PPG is highly correlated to daily motions and valuable for investigating practical PPG denoising and analysis applications for daily health monitoring. 
\bibliographystyle{IEEEtran}
\bibliography{ppg}
\end{document}